\documentclass[11pt]{article}
\usepackage{graphicx}
\usepackage{epsfig}

\newcommand{\BABARPubYear}    {04}

\newcommand{\BABARConfNumber} {29}
\newcommand{\SLACPubNumber} {10647}

\def\myprl  #1 #2 #3 {\jprl{#1},\ #2 (#3)}
\def\myplb  #1 #2 #3 {\plb{#1},\ #2 (#3)}
\def\myprd  #1 #2 #3 {\jprd{#1},\ #2 (#3)}
\def\mynim  #1 #2 #3 {\nim{#1},\ #2 (#3)}
\def\mypr   #1 #2 #3 {\pr{#1},\ #2 (#3)}
\def\mpp  {\ensuremath{m_{\pi\pi}}}
\def\ctheli  {\ensuremath{\cos\theta_h}}
\def\rhop  {\ensuremath{\rho(1450)}}

\input pubboard/babarsym

\long\def\inst#1{\par\nobreak\kern 4pt\nobreak
    {\it #1}\par\vskip 10pt plus 3pt minus 3pt}

\begin{document}
{\pagestyle{empty}
\begin{flushleft}
\end{flushleft}
\vskip 1cm

\begin{flushright}
\babar-CONF-\BABARPubYear/\BABARConfNumber \\
SLAC-PUB-\SLACPubNumber \\
August 2004 \\
\end{flushright}

\par\vskip 5cm

\begin{center}
\Large \bf Measurement of CP-violating parameters in fully reconstructed $B\to D^{(*)}\pi$ and $B\to D\rho$ decays.

\end{center}
\bigskip

\begin{center}
\large The \babar\ Collaboration\\
\mbox{ }\\
\today
\end{center}
\bigskip \bigskip

\begin{center}
\large \bf Abstract
\end{center}
We present a  preliminary measurement of  \CP-violating asymmetries in 
fully reconstructed $\Bz{\to}D^{(*)\pm}\pi^{\mp}$ and $\Bz{\to}D^{\pm}\rho^{\mp}$ decays in 
 approximately $110$ million \Y4S $\to$ \BB decays collected with the
\babar\ detector at the PEP-II asymmetric-energy $B$ factory at SLAC.
From a  maximum likelihood fit to the time-dependent decay distributions we obtain for the \CP-violating parameters:
$a^{D\pi} = -0.032\pm0.031 \,(\textrm{stat.})\pm 0.020 \,(\textrm{syst.}),   
c_{\rm lep}^{D\pi} = -0.059\pm0.055  \,(\textrm{stat.})\pm 0.033 \,(\textrm{syst.})$ on the $\Bz{\to}D^{\pm}\pi^{\mp}$ sample,
$a^{D^*\pi} = -0.049\pm0.031 \,(\textrm{stat.})\pm 0.020 \,(\textrm{syst.}),   
c_{\rm lep}^{D^*\pi} = +0.044\pm0.054  \,(\textrm{stat.})\pm 0.033 \,(\textrm{syst.})$ on the $\Bz{\to}D^{*\pm}\pi^{\mp}$ sample, and
$a^{D\rho} = -0.005\pm0.044 \,(\textrm{stat.})\pm 0.021 \,(\textrm{syst.}),   
c_{\rm lep}^{D\rho} = -0.147\pm0.074  \,(\textrm{stat.})\pm 0.035 \,(\textrm{syst.})$ on the $\Bz{\to}D^{\pm}\rho^{\mp}$ sample.

\vfill

\begin{center}

Submitted to the 32$^{\rm nd}$ International Conference on High-Energy Physics, ICHEP 04,\\
16 August---22 August 2004, Beijing, China

\end{center}

\vspace{1.0cm}
\begin{center}
{\em Stanford Linear Accelerator Center, Stanford University, 
Stanford, CA 94309} \\ \vspace{0.1cm}\hrule\vspace{0.1cm}
Work supported in part by Department of Energy contract DE-AC03-76SF00515.
\end{center}

\newpage
} 

\begin{center}
\small

The \babar\ Collaboration,
\bigskip

%
B.~Aubert,
R.~Barate,
D.~Boutigny,
F.~Couderc,
J.-M.~Gaillard,
A.~Hicheur,
Y.~Karyotakis,
J.~P.~Lees,
V.~Tisserand,
A.~Zghiche
\inst{Laboratoire de Physique des Particules, F-74941 Annecy-le-Vieux, France }
% ...

\end{center}\newpage

\section{INTRODUCTION}
\label{sec:Introduction}
While the measurement of $\sin 2\beta$ is now a precision measurement~\cite{babar_sin2b,belle_sin2b},
the constraints on the other two angles of the Unitarity Triangle~\cite{CKM}, $\alpha$ and $\gamma$, are still 
limited by statistics and/or by theoretical uncertainties.
This conference paper reports on the measurement of \CP-violating asymmetries
in  $\Bz{\to} D^{(*)\pm} \pi^{\mp}$ and  $\Bz{\to} D^{\pm} \rho^{\mp}$ 
decays~\cite{chconj} in $\FourS\to \BB$ decays, asymmetries
which are related to $|\sin(2\beta+\gamma)|$~\cite{sin2bg,fleischer}.
This analysis updates the results already published in ~\cite{olds2bg} 
by including  a new decay mode ($\Bz{\to} D^{\pm} \rho^{\mp}$) and 
a larger data sample (110 instead of 88 million \Y4S $\to$ \BB decays).

The time evolution of $B^0{\to} D^{(*)\pm} h^{\mp}$ decays, where h is a meson 
made of a $u$ and a $d$ quarks, is
sensitive to $\gamma$ because the CKM-favored 
decay $\Bzb{\to} D^{(*)+} h^-$, which amplitude is proportional to the CKM matrix
elements $V^{}_{cb}V^*_{ud}$, and the doubly-CKM-suppressed  
decay $\Bz{\to} D^{(*)+}h^-$, which amplitude is proportional to $V_{cd}V^*_{ub}$ interfere due  to the \Bz-\Bzb\ mixing. 
The relative weak phase between the two  amplitudes is $\gamma$,
and, when combined with $\BzBzb$ mixing, yields
a weak phase difference of $2\beta+\gamma$ between the interfering amplitudes.

The decay rate distribution for $B^0{\to} D^{\pm}h^{\mp}$ decays is
\begin{eqnarray}
f^{\pm}(\eta,\deltat) &=& \frac{e^{-\left|\deltat\right|/\tau}}{4\tau} \times [1 \mp S_\zeta \sin(\deltamd\deltat) \mp\eta C \cos(\deltamd\deltat)]\,,
\label{eq:fplus}
\end{eqnarray}
where $\tau$ is the \Bz lifetime, neglecting the decay width difference, $\deltamd$ is the $\Bz\Bzb$ mixing frequency,
and $\deltat = t_{\rm rec} - \t_{\rm tag}$ is the time of the  
$B^0\to D^{\pm} \pi^{\mp}$ decay ($B_{\rm rec}$) relative to the decay of the other $B$ ($B_{\rm tag}$). 
In this equation the upper (lower) sign refers to the flavor of $B_{\rm tag}$ as \Bz(\Bzb),
while $\eta=+1$ ($-1$) and $\zeta=+$ ($-$) for the final state $D^{-}h^{+}$ ($D^{+}h^{-}$).
In the Standard Model, the $S$ and $C$ parameters can be expressed as 
\begin{eqnarray}
S_\pm = -\frac{2\textrm{Im}(\lambda_\pm)}{1+|\lambda_\pm|^2}\,, \hspace{0.4cm} {\rm and} 
\hspace{0.4cm}
C=\frac{1-r^2}{1+r^2}\,,
\label{eq:cands}
\end{eqnarray}
where  $r\equiv |\lambda_+| = 1/|\lambda_-|$ and
\begin{eqnarray}
\label{eq:lambda}
 \lambda_\pm =\frac{q}{p}
 A(\Bzb{\to} D^{\mp}\pi^\pm)/A(\Bz{\to}D^{\mp}\pi^\pm)
=r^{\pm 1}e^{-i(2\beta+\gamma\mp\delta)}.
\end{eqnarray}
 Here
$\frac{q}{p}$ is a function of the elements of the mixing
matrix~\cite{PDG}, and $\delta$ is the relative strong phase
between the
two contributing amplitudes.
In these equations the parameters $r$ and $\delta$ depend on the choice of 
the final state and will be indicated as  $r^{D\pi}$, $\delta^{D\pi}$, in the  $\Bz{\to}D^{\pm}\pi^{\mp}$ case,
  $r^{D^*\pi}$, $\delta^{D^*\pi}$, in the  $\Bz{\to}D^{*\pm}\pi^{\mp}$ case
\footnote{According to Ref.~\cite{fleischer} the strong phase is actually $\delta^{D^*\pi}+\pi$, but this does not
affect this measurement. },
and   $r^{D\rho}$, $\delta^{D\rho}$, in the  $\Bz{\to}D^{\pm}\rho^{\mp}$ case. 

Interpreting the  $S$ and $C$ parameters 
 in terms of the angles of the Unitarity Triangle requires the 
measurement of the $r$
parameters as detailed in Sec.~\ref{sec:analysis}. Since the amplitude in the numerator of Eq.~\ref{eq:lambda} is  suppressed
with respect to the one in the denominator, the $r$ parameters are expected
to be small ($\sim 0.02$) and they cannot therefore be extracted from the measurement
of $C$ with the current statistics. They can be determined, 
assuming $SU(3)$ symmetry and neglecting contributions from annihilation diagrams,
 from the ratios of branching fractions  $\BR(\Bz{\to} D_s^{(*)+}\pi^-)/{\BR(\Bz{\to}
  D^{(*)-}\pi^+)}$ and $\BR(\Bz{\to} D_s^{(*)+}\rho^-)/{\BR(\Bz{\to}
  D^{(*)-}\rho^+)}$~\cite{sin2bg,olds2bg,dsrho}. Since there is no evidence yet of 
$\Bz{\to} D_s^{(*)+}\rho^-$ decays, 
 we will report here the result of the
measurement of  CP-violating parameters defined in Eq.~\ref{completepdf}, but will not be able to include the interpretation in
terms of $\sin(2\beta+\gamma)$. Nonetheless, the addition of the $\Bz\to D^{\pm}\rho^{\mp}$ mode to the analysis is of interest because on one side
unexpectedly large values for the CP-violating parameters would signal new physics and on the other side when the statistics will be
sufficient it will allow one more measurement of  $\sin(2\beta+\gamma)$, redundant with respect to the one made with the other modes.

\section{ ANALYSIS OVERVIEW }
\label{sec:analysis}
This measurement is based on  110 million \Y4S $\to$ \BB decays
collected with the \babar\ detector~\cite{detector} at the PEP-II asymmetric-energy $B$ factory at SLAC.
We use Monte Carlo simulation of the \babar\ detector based on
GEANT4~\cite{geant} to validate the analysis procedure and to estimate
some of the backgrounds.
The analysis strategy is identical to our previous publication
on this topic~\cite{olds2bg}.

Candidate $\Bz\to D^{(*)\pm}\pi^{\mp}$ and  $\Bz\to D^{\pm}\rho^{\mp}$
 decays are reconstructed with the \Dstarp decaying
to $\Dz\pip$, where the \Dz subsequently decays to one of the four modes
$K^{-}\pi^{+}$, $K^{-}\pi^{+}\piz$, $K^{-}\pi^{+}\pi^{-}\pi^{+}$, or $\KS\pi^{+}\pi^{-}$,
the \Dp decaying into $K^{-}\pi^{+}\pi^{+}$ and $\KS\pi^{+}$, and the $\rho^+$ decaying into $\pip\piz$.
The $D^{(*)}$ candidates are then combined with either a single track or 
a track and a \piz candidate with invariant mass in the $\rho$ window,
$620<\mpp<920\mevcc$. Finally, exploiting the spin properties of
the decay of a pseudo-scalar meson into another pseudo-scalar and a vector, we require  the 
cosine of the angle $\theta_l$ between the charged pion and the $\rho$ candidate in the $\rho$ rest frame 
 to satisfy $|\ctheli|>0.4$.

Signal and background are discriminated by two kinematic variables:
the beam-energy substituted mass, $\mes \equiv \sqrt{(\sqrt{s}/2)^{2} - {p_B^*}^2}$,
and the difference between the $B$ candidate's measured energy and the beam energy, 
$\DeltaE \equiv E_{B}^* - (\sqrt{s}/2)$.
$E_{B}^*$ ($p_B^*$) is the energy (momentum) of the \B\ candidate
in the $e^{+}e^{-}$ center-of-mass frame, and $\sqrt{s}$ is the total center-of-mass energy.
The signal region is defined to be $|\DeltaE| <3\sigma$, where the resolution $\sigma$ is mode-dependent
and is approximately 20\mev as determined from data. 
\begin{figure}[!tbp]
\begin{center}
\epsfig{figure=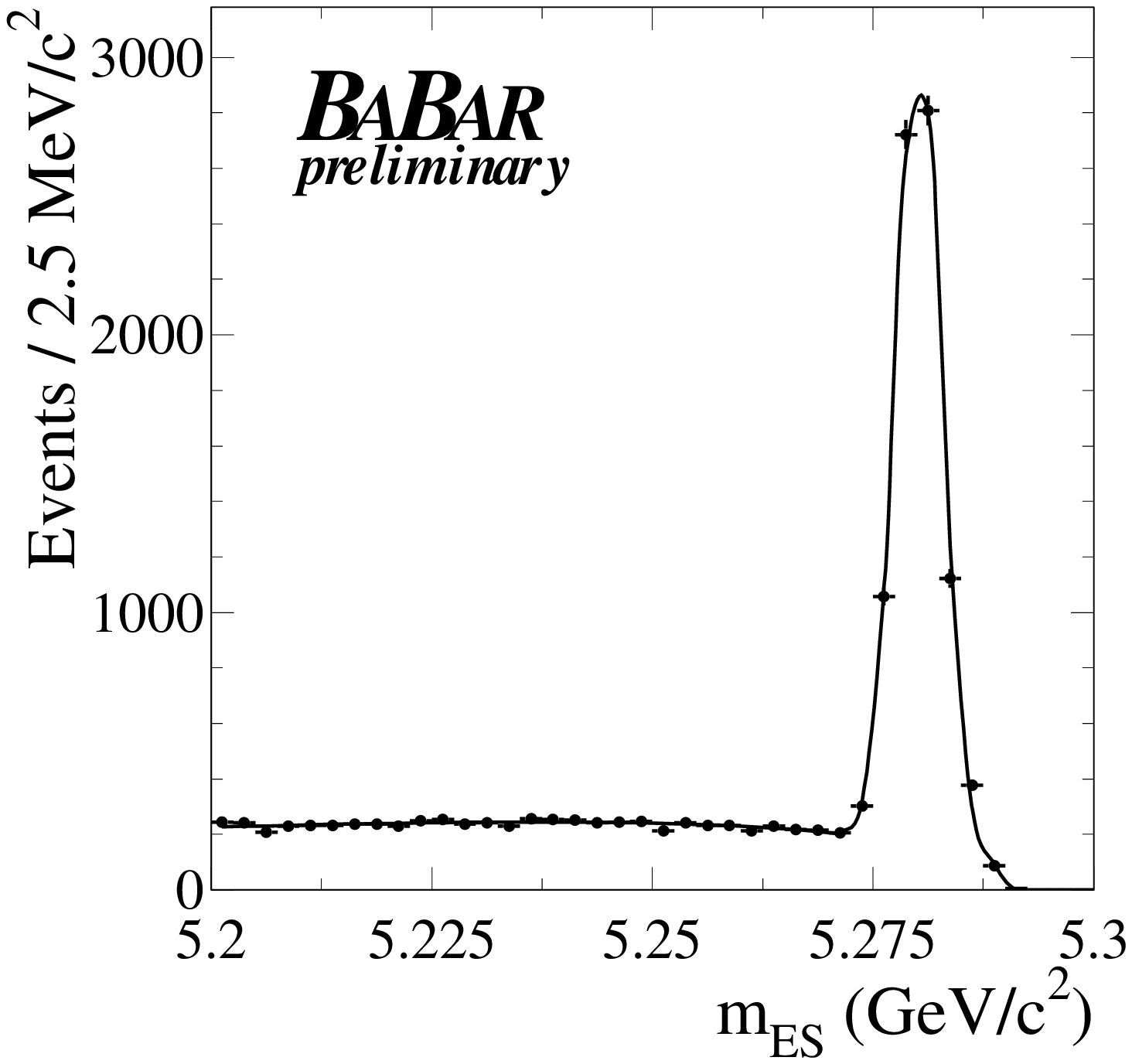,width=5.4cm}
\epsfig{figure=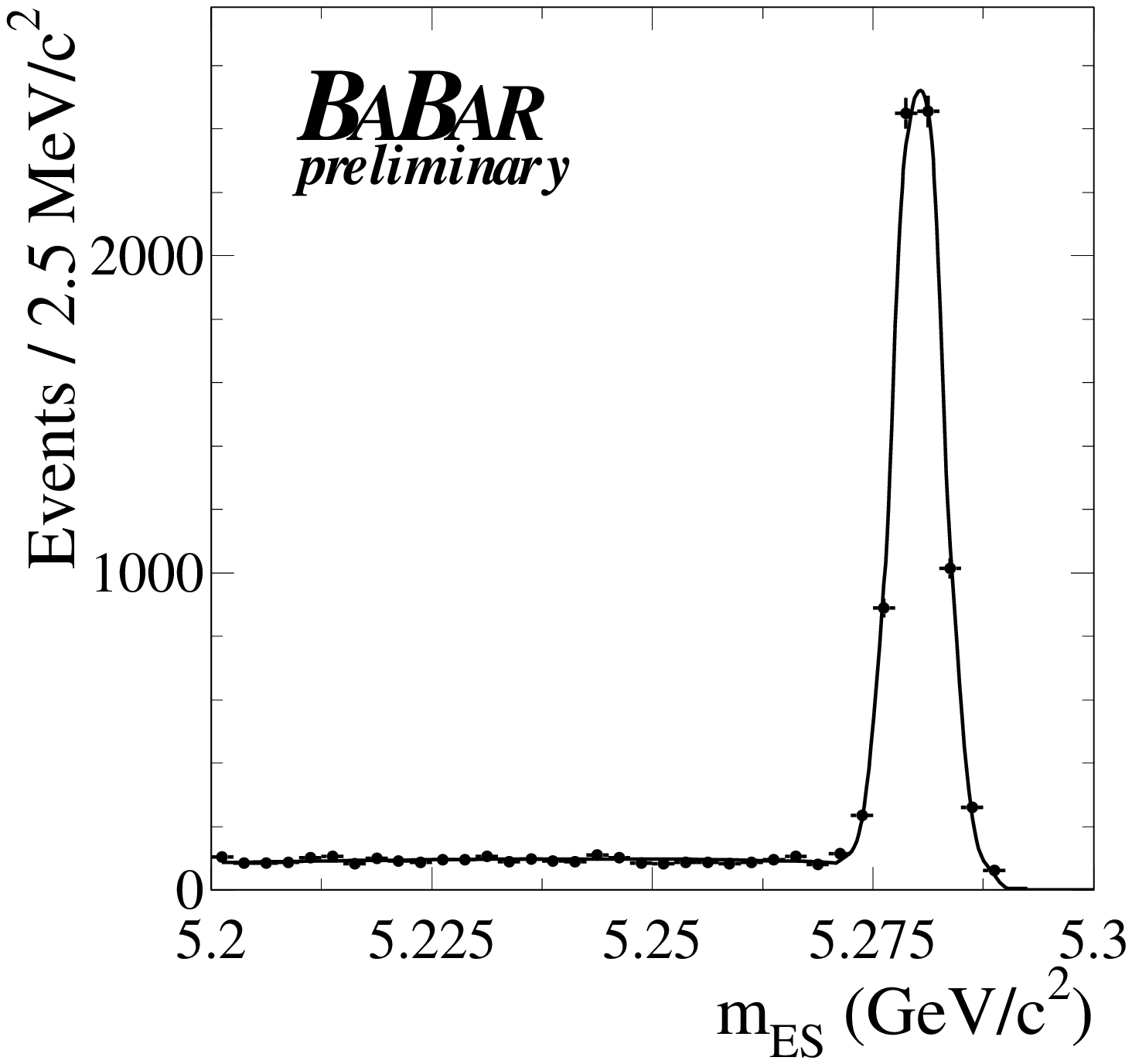,width=5.4cm}
\epsfig{figure=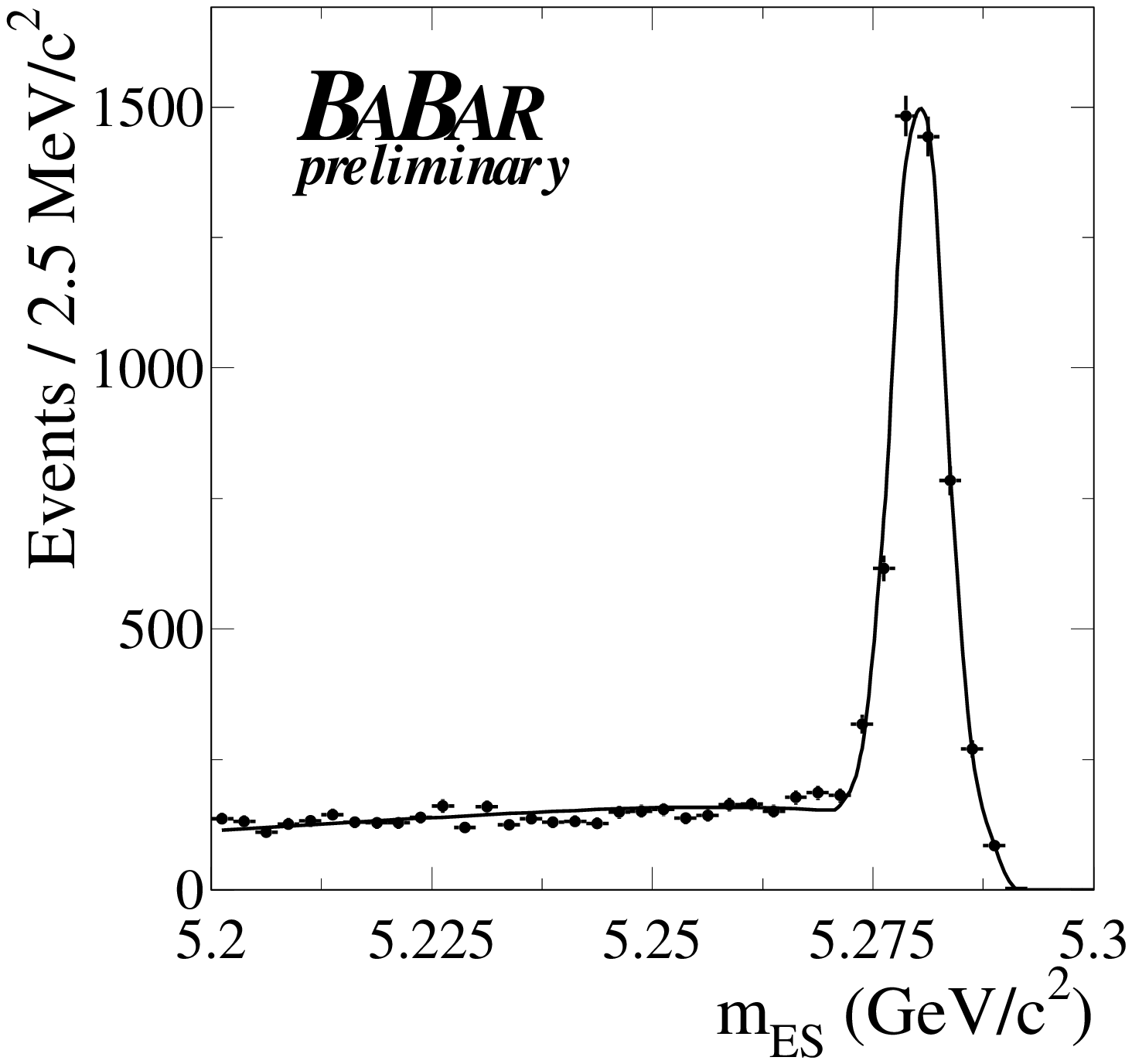,width=5.4cm}
\end{center}
\caption{\mes\ distributions in the \DeltaE\ signal region for, from left to right, the $\Bz
  \rightarrow D^{\pm}\pi^{\mp}$, $B^0 \rightarrow D^{*\pm}\pi^{\mp}$
and $B^0 \rightarrow D^{\pm}\rho^{\mp}$ sample for events with tagging information. A fit to a Gaussian plus a threshold is overlaid.}
\label{mes}
\end{figure}
Figure~\ref{mes} shows the $\mes$ distribution for candidates in the signal region. 

To identify the flavor of $B_{\rm tag}$, each event
is assigned by a neural network to one of four hierarchical, mutually exclusive
tagging categories:
one lepton and two kaon categories based  on the charges of 
identified leptons and kaons, and a fourth category 
for remaining events~\cite{babar_sin2b}.
The effective tagging efficiency is ($28.1\pm 0.7$)\%. The time
difference $\deltat$ is calculated from the measured separation along the
beam collision axis, $\deltaz$, between the 
reconstructed ($B_{\rm rec}$) and tagged ($B_{\rm tag}$) decay vertexes. We determine the  $B_{\rm rec}$ vertex from its charged tracks. 
The  $B_{\rm tag}$ decay vertex is obtained by fitting tracks that do not belong
to $B_{\rm rec}$, 
imposing constraints from the $B_{\rm rec}$ momentum and the beam-spot location. 
The $\deltat$ resolution is approximately 1.1 ps.

An unbinned likelihood fit is performed to the time distribution of events in this sample. The likelihood accounts 
 for possible \CP\ violation on the tag side~\cite{DCSD}:
for each tagging category $i$ and for each decay mode $\mu=D^{\mp}\pi^\pm, D^{*\mp}\pi^\pm,
D^{\mp}\rho^\pm$
\begin{eqnarray}
f^{\pm{\mu}}_i(\eta,\deltat) &=&
  \frac{e^{-\left|\deltat\right|/\tau}}{4\tau} \times [ 1 \mp (a^{\mu}
  \mp \eta b_i - \eta c_i^{\mu})\nonumber \\
  && \sin(\deltamd\deltat)\mp\eta\cos(\deltamd\deltat)]\,,
\label{completepdf}
\end{eqnarray}
where in the Standard Model
\begin{eqnarray}\nonumber
&a^{\mu}&=\ 2r^{\mu}\sin(2 \beta+\gamma)\cos\delta^{\mu}\,, \\ \nonumber
&b_i&=\ 2r^\prime_i\sin(2 \beta+\gamma)\cos\delta^\prime_i\,, \\
&c_i^{\mu}&=\ 2\cos(2 \beta+\gamma) (r^{\mu}\sin\delta^{\mu}-r^\prime_i\sin\delta^\prime_i)\,.
\label{acdep}
\end{eqnarray}
Here $r^\prime_i$ ($\delta^\prime_i$) is, for each tagging category, the effective amplitude (phase) 
used to parameterize the tag side interference.  
Terms of order $r^{\mu\ 2}$ and $r^{\prime 2}_i$ have been neglected.
Results are quoted only  for the six $a^\mu$ and $c^\mu_{lep}$ parameters,
 which are independent of the unknowns 
$r^\prime_i$ and $\delta^\prime_i$ (semileptonic B decays have no doubly CKM-
suppressed  contributions and therefore $r^\prime_{lep}$=0).
The other parameters are constrained by 
the fit, but, as they depend 
on $r^\prime_i$ and
$\delta^\prime_i$, 
they do not contribute to the interpretation of the result in terms 
of $\sin(2\beta+\gamma)$.

\section{ SAMPLE COMPOSITION}
\label{samplecompo}
The background can be separated in two categories, one of which is due to random combinations
of particles in the event (${combinatorial }$ background) and the other is due to $B$ decays 
into similar final states, which therefore has the \mes\ distribution similar to the signal
(${peaking}$ background).
To separate the combinatorial background, 
 the \mes\ distribution is fit with the sum of a threshold function~\cite{Argus}
 and a Gaussian, with a width of about 2.5\mevcc, to describe the signal. 
\begin{table}[ht]
\begin{center}
\caption{Yields, fraction of combinatorial background $f_{comb}$ and of peaking backgrounds $f_{peak}$ of the selected samples.\label{tab:yields}}
\begin{tabular}{|l|c|c|c|c|}\hline
Decay  & yields & $f_{comb}(\%)$ & \multicolumn{2}{|c|}{$f_{peak}$(\%)} \\ 
mode &  & &  \Bz & \Bpm\\  
\hline
$\Bz{\to}D^{\pm}\pi^{\mp}$  &$7611\pm$97& 8.6&0.21$\pm$0.06&0.93 $\pm$ 0.23 \\
$\Bz{\to}D^{*\pm}\pi^{\mp}$ &$7068\pm$89& 4.2&0.13$\pm$0.06&0.93 $\pm$ 0.10 \\
$\Bz{\to}D^{\pm}\rho^{\mp}$ &$4400\pm$79&12.7& -0.01$\pm$0.07 & 0.31 
$\pm$ 0.13 \\
\hline
\end{tabular}
\end{center}
\end{table}

We considered all possible sources of background 
 peaking in the \mes\ signal region,
those coming from decays into open-charm final states
similar to that of the
signal  (e.g.,  $\Bm\to\Dstarz\pim,\rho^-$ or
$\Bzb\to\Dstarp\pim,\rho^-$),
 and those arising from charmless decays   into the same
final state as the signal. We estimated their contributions on MC simulation, 
varying the branching fractions within errors when observed and within the 
existing upper limits otherwise. The Gaussian yields and the amount of peaking background are summarized in
Table ~\ref{tab:yields}, identified by the source,either neutral or charged $B$ mesons.
                                                                                
In the case of the $B^0 \rightarrow D^{*\pm}\rho^{\mp}$ decays, an additional source of 
background must be considered, which has the same final state,
 $B^0 \rightarrow D^{\pm}\pi^{\mp}\piz$, where the $\pi^{\mp}\piz$ system is not
produced by a $\rho$ meson. This background component can be studied by looking at the 
distribution of \mpp\ and $\ctheli$. 
When the $\pi^{\mp}$ and $\piz$ come from a $\rho$ meson, 
\mpp\ follows the $\rho$ lineshape, while  $\ctheli$ is distributed 
as $(\ctheli)^2$. In order to satisfy the Bose-Einstein statistics,  $\pi^{\mp}$ and $\piz$  must 
come from a resonance with odd spins ($J=1,3,...$), and therefore, to first approximation,
we only consider excited states of the $\rho$ meson. 
In the mass range of interest, $\rhop$ is the only possible candidate.
For $B^0 \rightarrow D^{\pm}\rhop$ decays, the \mpp\ distribution
would be peaked at higher masses (the \rhop\ pole mass is $(1465\pm25)\mevcc$, 
its width is $(400\pm60)$~\cite{PDG}), but would have the same $(\ctheli)^2$ distribution.
In order to have a  \ctheli\ distribution different from the signal, the background
would have to come from non-resonant  $B^0 \rightarrow D^{\pm}\pi^{\mp}\piz$ decays.
We therefore consider the possibility of having three components: the signal, 
$B^0 \rightarrow D^{\pm}\rho^{\mp}$,  $B^0 \rightarrow D^{\pm}\rhop$, and an S-wave 
non-resonant component, $B^0 \rightarrow D^{\pm}(\pi^{\mp}\piz)_{nr}$.

The presence of these two additional components  would imply that the value  of \\
$A(\Bzb{\to} D^{\mp}\pi^\pm\piz)/A(\Bz{\to}D^{\mp}\pi^\pm\piz)$ 
 (and thus $\lambda^{D\rho}$, see Eq.~\ref{eq:lambda})
is not  constant with \mpp:
\begin{eqnarray}
&&A(\Bzb{\to} D^{\mp}\pi^\pm\piz)/A(\Bz{\to}D^{\mp}\pi^\pm\piz)=\\ \nonumber
& & \frac{A(\bar{B^0} \to D^-\rho+)A_{\rho}(\mpp)
+A(\bar{B^0} \to D^-\rhop^+)A_{\rhop}(\mpp)
+A(\bar{B^0} \to D^-\pi^+\pi^0)A_{nr}(\mpp)}
{A(B^0 \to D^-\rho^+)A_{\rho}(\mpp)
+A(B^0 \to D^-\rhop^+)A_{\rhop}  (\mpp)
+A(B^0 \to D^-\pi^+\pi^0)A_{nr}(\mpp)}.
\label{lambda1} 
\end{eqnarray}
Here  $A_{\rho}(\mpp), A_{\rhop}(\mpp)$, and $A_{nr}(\mpp)$ are the amplitudes of the 
three components and depend on \mpp
\footnote{In a Dalitz analysis they would depend also on the $D$ $\pi$ invariant mass, 
but here we integrate over it to
study a single variable at a time.}. 

\begin{figure}[!tbp]
\begin{center}
\epsfig{figure=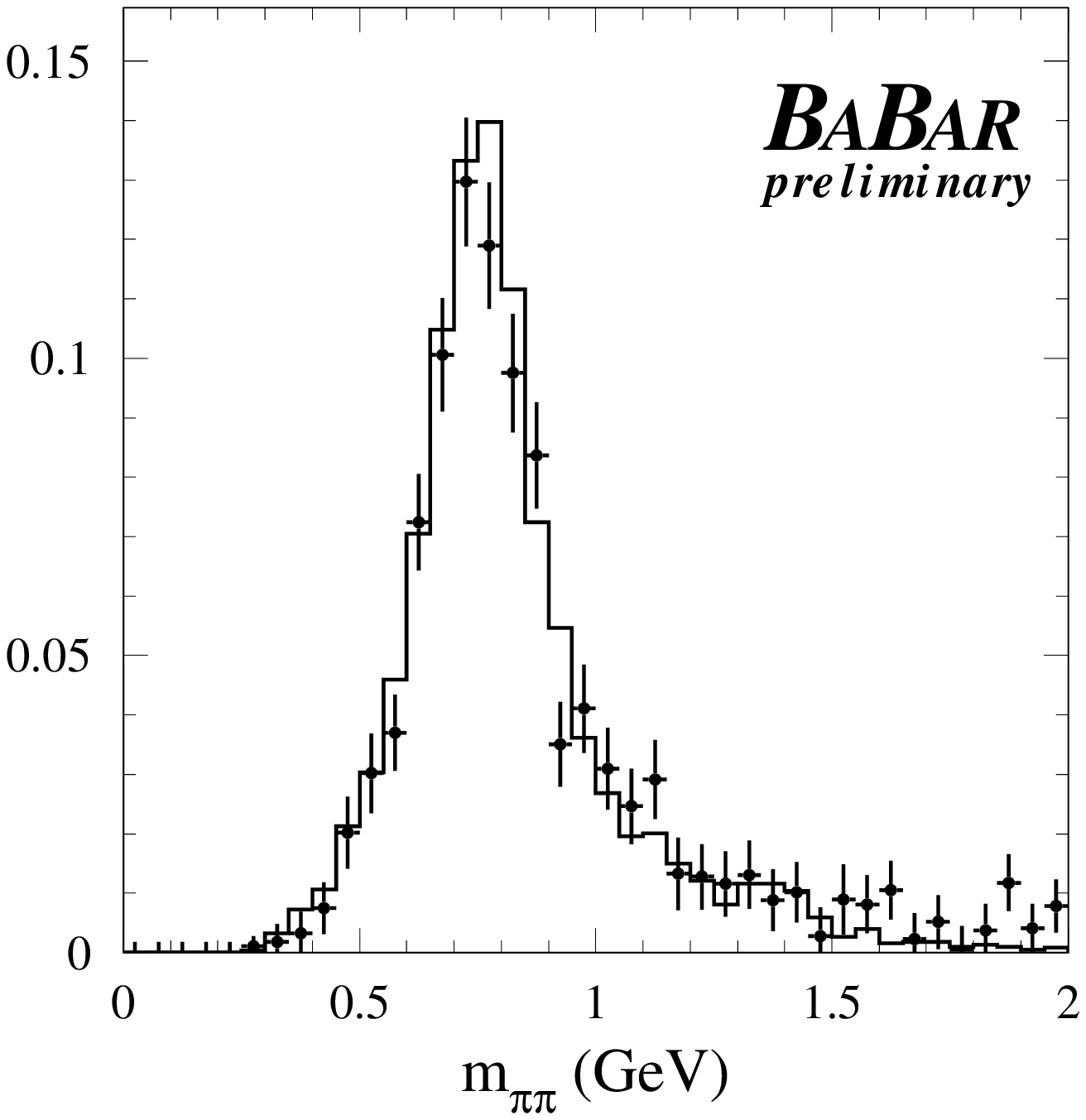,width=0.45\linewidth,clip=}
\epsfig{figure=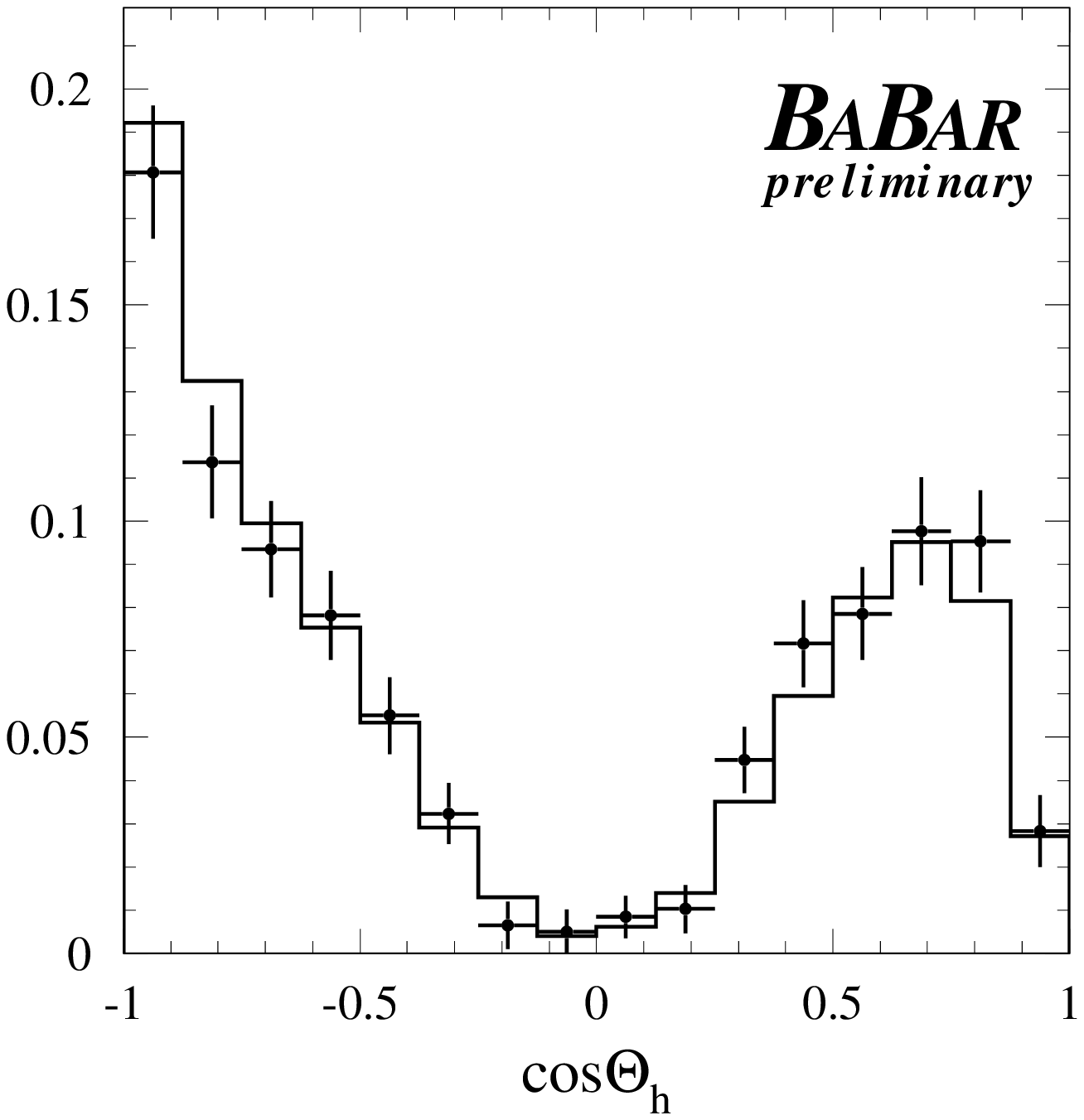,width=0.45\linewidth,clip=}
\end{center}
\caption{Sideband-subtracted \mpp \ and \ctheli \ distributions on data (dots) 
and  $\Bz{\to} D^{\pm} \rho^{\mp}$  simulation  (open histogram). }
\label{datamc_dalitz}
\end{figure}

Figure~\ref{datamc_dalitz} shows the comparison of  the \mpp\ and \ctheli\ distributions
between data and a simulation of pure $\Bz{\to} D^{\pm} \rho^{\mp}$ decays.
The good agreement suggests that the contribution of higher resonances is negligible. To quantify this statement, we perform a fit 
to the \mpp\ distribution including the lineshapes of the three components, and extract the relative amplitudes and phases. Given this set of 
amplitudes and phases, we generate the \mpp\ distribution and, for each value of \mpp, the appropriate \deltat\ distribution. We then fit these 
samples with the likelihood described in Eq.(\ref{completepdf}), which ignores the dependence of $\lambda^{D\rho}$  on \mpp. We repeat this procedure 
for several sets of amplitudes and phases according to the measured covariance matrix and find the bias induced to the $a$ ($c$) parameters to be at most
0.0018 (0.0047). This maximum bias will be included as a systematic error.

An unbinned maximum likelihood fit is performed on the selected $B$ candidates using the signal \deltat\ distribution
in Eq.(~\ref{completepdf}), convoluted with a three-Gaussian 
resolution function.
The  probabilities of incorrect tagging ($w_i$) are accounted for by
multiplying the $a^{\mu}$, $c_i^{\mu}$
 parameters and the $\cos(\deltamd\deltat)$
term by the dilutions $D_i=1-2w_i$.
The resolution function and the tagging parameters 
are consistent within errors with previous \babar\
 analyses~\cite{babar_sin2b}.

The combinatorial background is parametrized as the sum of a component
with zero lifetime and one with an effective lifetime fixed to the value obtained from simulation.
The fraction of each background component is determined from the events in the \mes\ sidebands, $5.2<\mes<5.27\gevcc$,
while the \deltat\ resolution is a double-Gaussian fitted on  the same data as the CP-violating parameters. The charmed background coming from $B^{\pm}$ mesons
is modeled by an exponential with the  $B^{\pm}$ lifetime, and
its amount is fixed to the value predicted by simulation.
The charmed and charmless backgrounds from $B^{0}$
mesons are neglected in the nominal fit, but are considered in evaluating the systematic uncertainties.

\section{RESULTS}\label{sec:Results}
From the unbinned maximum likelihood fit we obtain:
\begin{eqnarray*}
\begin{array}{rclcrcl}
a^{D\pi}&=&-0.032 \pm 0.031 \ (\mbox{stat.}) &,&
c_{lep}^{D\pi}&=&-0.059 \pm 0.055 \ (\mbox{stat.}) \\ \nonumber
a^{D^*\pi}&=&-0.049 \pm 0.031 \ (\mbox{stat.})&,&
c_{lep}^{D^*\pi}&=&0.044 \pm 0.054 \ (\mbox{stat.}) \\ \nonumber
a^{D\rho}&=&-0.005 \pm 0.044 \ (\mbox{stat.}) &,&
c_{lep}^{D\rho}&=&-0.147 \pm 0.074 \ (\mbox{stat.}) \\ \nonumber
\end{array}
\end{eqnarray*}

Table~\ref{tab:sys} shows the contributions to the systematic
uncertainty. Errors are found to be  independent of the
$B_{rec}$ reconstruction mode because the systematic
effects are dominated by uncertainties in the $B_{tag}$ reconstruction. In the 
table we compare the results obtained with $\Bz{\to}D^{(*)\pm}\pi^{\mp}$ decays and  
$\Bz{\to}D^{\pm}\rho^{\mp}$.

\begin{table}[ht]
\begin{center}
\caption{Systematic uncertainties on the $a$ and $c$ parameters.\label{tab:sys}}
\begin{tabular}{|l|l|l|l|l|}\hline
  & \multicolumn{2}{|c|}{ $\Bz{\to}D^{(*)\pm}\pi^{\mp}$}& \multicolumn{2}{|c|}{$\Bz{\to}D^{\pm}\rho^{\mp}$}\\ \hline
Source & $\sigma_{a}$ & $\sigma_{c}$ & $\sigma_{a}$ & $\sigma_{c}$  \\
\hline
Vertexing ($\sigma_{\Delta t}$) & 0.015 & 0.026&0.017&0.031\\
Fit ($\sigma_{\rm fit}$) & 0.011 & 0.019&0.009&0.016\\
Model ($\sigma_{\rm mod}$) & 0.006 & 0.007&0.0007&0.0015\\
Tagging ($\sigma_{\rm tag}$) & 0.004 & 0.0034&0.0028&0.0033\\
Background ($\sigma_{\rm bkg}$) & 0.0012 & 0.0027&0.006&0.0031\\
Dependence from $m_{\pi\pi^0}$ ($\sigma_{\rm \lambda dep}$) &  & &0.0018&0.0047\\
\hline
Total ($\sigma_{\rm tot}$) & 0.020 & 0.033&0.021&0.035\\ \hline
\end{tabular}
\end{center}
\end{table}

The impact of a possible mismeasurement of $\deltat$ ($\sigma_{\Delta t}$)
has been estimated by comparing the  different
parameterizations of  the resolution function, varying
the position of the beam spot, and
 the absolute $z$ scale   within their uncertainties, 
 and loosening and tightening the 
quality criteria on the reconstructed vertex. We also estimate the
impact of the uncertainties on the alignment of the silicon detector (SVT)
by repeating the measurement on simulated events, intentionally
misaligning the SVT in the simulation.
As systematic uncertainty of the fit ($\sigma_{\rm fit}$), 
 we quote the upper limit on the bias on the $a^{\mu}$ and $c^{\mu}$,  
 as estimated
from samples of fully simulated events. The model error
($\sigma_{\rm mod}$) contains the uncertainty on the $B^{0}$ lifetime and \deltamd,
varied by the uncertainties on the world averages~\cite{PDG}, and the impact
of neglecting higher order terms in $r$ or $r^\prime_i$ in Eq.(\ref{completepdf}). 
The tagging error ($\sigma_{\rm tag}$)
 is estimated
considering possible differences in tagging efficiency between \Bz\ and \Bzb\ and 
allowing for different $\Delta t$ resolutions for correctly and incorrectly tagged events. 
We also account for uncertainties on the background ($\sigma_{\rm bkg}$)
 by varying the effective lifetimes, dilutions, \mes\
shape parameters, signal fractions, 
and background \CP\ asymmetry up to five times the expected \CP\ asymmetry for signal.
For the  $B\to D\rho$  decay we also include the maximum bias of the $a$ and $c_{lep}$ parameters due
 to the possible dependence of $\lambda$ on  the $\pi \pi^0$ invariant mass ($\sigma_{\rm \lambda dep}$), 
as discussed in section \ref{samplecompo}.

\section{CONCLUSIONS}
We studied the time evolution of fully reconstructed
$\Bz{\to}D^{(*)\pm}\pi^{\mp}$ and $\Bz{\to}D^{\pm}\rho^{\mp}$ decays
in a data sample of 110 million \Y4S $\to$ \BB decays.
\CP-violation arising from the interference of the CKM-suppressed and the CKM-favored amplitudes 
 is expected to be small but sensitive to $\sin(2\beta+\gamma)$.

The \CP-violating parameters defined in Eq.~\ref{acdep} are measured  to be
\begin{eqnarray*}
\begin{array}{rclcrcl}
a^{D\pi}&=&-0.032 \pm 0.031 \ (\mbox{stat.})\pm 0.020 \ (\mbox{syst.})&,&
c_{lep}^{D\pi}&=&-0.059 \pm 0.055 \ (\mbox{stat.})\pm 0.033 \ (\mbox{syst.})\\ \nonumber
a^{D^*\pi}&=&-0.049 \pm 0.031 \ (\mbox{stat.})\pm 0.020 \ (\mbox{syst.})&,&
c_{lep}^{D^*\pi}&=&+0.044 \pm 0.054 \ (\mbox{stat.})\pm 0.033 \ (\mbox{syst.})\\  \nonumber
a^{D\rho}&=&-0.005 \pm 0.044 \ (\mbox{stat.})\pm 0.021 \ (\mbox{syst.})&,&
c_{lep}^{D\rho}&=&-0.147 \pm 0.074 \ (\mbox{stat.})\pm 0.035 \ (\mbox{syst.}).  \nonumber
\end{array}
\end{eqnarray*}
No significant CP asymmetry is observed thus far.

\section{ACKNOWLEDGMENTS}

We are grateful for the 
extraordinary contributions of our \pep2\ colleagues in
achieving the excellent luminosity and machine conditions
that have made this work possible.
The success of this project also relies critically on the 
expertise and dedication of the computing organizations that 
support \babar.
The collaborating institutions wish to thank 
SLAC for its support and the kind hospitality extended to them. 
This work is supported by the
US Department of Energy
and National Science Foundation, the
Natural Sciences and Engineering Research Council (Canada),
Institute of High Energy Physics (China), the
Commissariat \`a l'Energie Atomique and
Institut National de Physique Nucl\'eaire et de Physique des Particules
(France), the
Bundesministerium f\"ur Bildung und Forschung and
Deutsche Forschungsgemeinschaft
(Germany), the
Istituto Nazionale di Fisica Nucleare (Italy),
the Foundation for Fundamental Research on Matter (The Netherlands),
the Research Council of Norway, the
Ministry of Science and Technology of the Russian Federation, and the
Particle Physics and Astronomy Research Council (United Kingdom). 
Individuals have received support from 
CONACyT (Mexico),
the A. P. Sloan Foundation, 
the Research Corporation,
and the Alexander von Humboldt Foundation.


\begin{thebibliography}{99}
\bibitem{chconj}
Charge conjugation is implied in this paper, unless 
explicitly stated otherwise.
\end{thebibliography}
\end{document}